# EARLY STAKEHOLDER ENGAGEMENT FOR A POSSIBLE NEW MULTIPURPOSE RESEARCH REACTOR FOR CANADA

Z. YAMANI, L. WALTERS, A. SIDDIQUI
*Canadian Nuclear Laboratories*
*Chalk River, ON K0J 1J0, Canada*

K. HUYNH
*Atomic Energy Canada Limited, Ottawa, ON K1P 5G8, Canada*

## ABSTRACT

Canada has a rich history of nuclear technology development. Since the 1940s, nuclear research infrastructure and facilities, such as National Research Universal (NRU) reactor at Canadian Nuclear Laboratories in Chalk River, Ontario, have played a key role for R&D and for building Canadian expertise and competency in nuclear technology. The NRU reactor retired in 2018. Since the needs of stakeholders vary with time, consideration of a new multipurpose research reactor for Canada must contemplate current user input for their requirements. As an early step in such consideration, a systematic approach was employed to engage various national stakeholders from academia, industry, and research organizations, including the government. Several virtual workshops were held, each with a specific theme around utilizations. To provide international context, our workshops also included presentations from leading nuclear laboratories around the world. We provide a summary of Canada's approach and the main findings of this early stakeholder engagement.

1. Introduction

Canada's only major high-flux multipurpose research reactor, the NRU reactor, retired in 2018 [1]. For over 60 years, NRU was a pillar of nuclear research and development (R&D) in Canada [2-6]. It enabled innovation in many areas, such as nuclear power, materials and fuel testing, neutron beam research, radioisotope production, and nuclear medicine. The work conducted at NRU contributed to medical treatments for well over 500 million patients globally in its lifetime [4]. It also led to a Nobel Prize in Physics [7] and the demonstration of online refuelling of natural uranium fuel moderated and cooled with heavy water—the distinguishing characteristics of CANDU (CANada Deuterium Uranium) power reactors [8, 9], which are at the heart of Canada's internationally respected nuclear sector, which includes a $17 billion/year domestic nuclear industry [10].

Over the past decades, there have been studies and assessments on how a new research reactor could add value to science and technology (S&T) in Canada (for example see Refs. [11, 12]). With rapid developments in nuclear technology, renewed interest in nuclear as a low-carbon-emitting power source, and as Canada embarks on the small modular reactors (SMRs) deployment initiatives, and recent developments in nuclear power generation in Canada [13-16] and a surge of interest in nuclear power globally [17], it is time to review the potential needs, opportunities, challenges, and benefits of a new high-flux multipurpose research reactor for Canada.

As the first step in the consideration of a new research reactor and to facilitate a discussion among experts from various Canadian organizations, under the support of Atomic Energy

Limited (AECL) [18], Canadian Nuclear Laboratories (CNL) [19] organized a series of workshops from March to July 2021. The workshops provided a broad overview and context from a variety of stakeholders in academy, industry, government, and experts from key national organizations to present and discuss the needs and requirements of their respective communities for a new research reactor in Canada, as well as the challenges that their communities face in performing required R&D activities without a domestic facility. In this paper, we provide a summary of our engagement approach and the main findings of the workshop series.

## 2. Engagement Approach and Details

The workshop series were developed by CNL under the guidance of AECL and ran as virtual events from March to July 2021. The main objectives of this stakeholder engagement were to stimulate discussion with stakeholders on the current and anticipated needs, opportunities, challenges, and benefits that should form part of the consideration of a new research reactor in Canada.

The first session was planned to establish context through a broad overview of research reactor utilization and the status of existing and upcoming facilities from both national and international perspectives. Four additional workshop sessions were held, each focusing on a key reactor utilization, to assess the technical needs and requirements of the stakeholders in more detail:

- Neutron beam research (April 27, 2021),
- Materials & fuel irradiation (May 17, 2021),
- Production of radioisotopes (June 21, 2021), and
- Safety & security (July 19, 2021).

All the workshop sessions were held virtually to maximize the participation of a variety of stakeholders and followed a similar format: short, invited presentations by experts from key Canadian and international institutions, followed by open discussion periods providing an interactive forum for free exchange of opinions pertinent to the needs and requirements for a new research reactor in Canada. A diverse range of reactor utilizations and considerations were covered in the workshop series, including nuclear energy generation, regulatory requirements, nuclear safety & security, training and development of highly qualified personnel (HQP) & skilled workers, clean energy (hydrogen technology) and the environment, radioisotopes (medical and non-medical), industrial processing & manufacturing, materials science and engineering, biotechnology, food security & agriculture, as well as basic science.

The workshop series attracted a lot of interest, with more than 150 individuals from a broad range of stakeholders from key organizations (nuclear power industry, academia, nuclear medicine, nuclear regulator, and federal and provincial government offices) attending the sessions and taking part in the discussions. The diversity of the participants' technical backgrounds and organizations ensured that a variety of perspectives were captured in this initial early stakeholder engagement process. Figure 1 provides highlights of the workshop series.

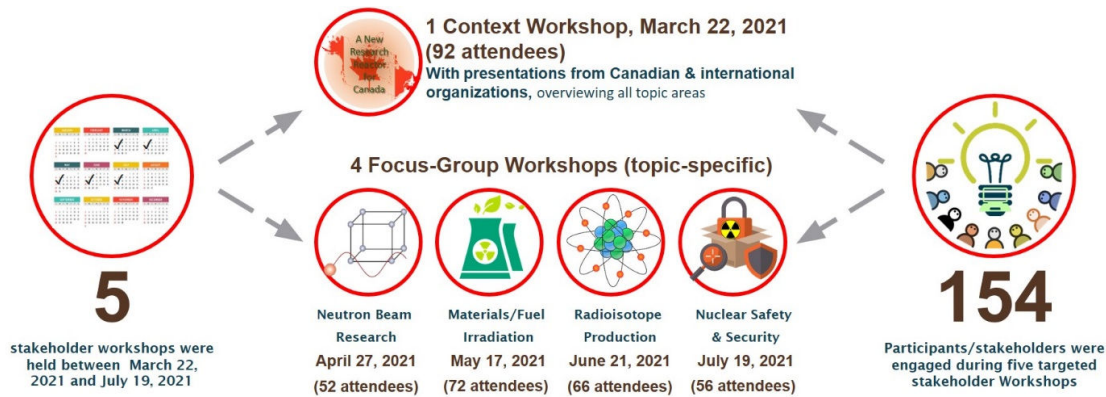

**Figure 1. Stakeholder engagement highlights. 154 individuals participated in the workshop series.**

This workshop series was funded by AECL's Federal Nuclear Science & Technology (FNST) Work Plan.

## 3. Key Observations and the Role of a New Research Reactor

### 3.1 Capability gaps in existing facilities

The workshop series revealed strong support among the participants that a new domestic source underpins the needs of various stakeholders and communities across the Canadian landscape that cannot be fully and easily met through current and planned Canadian facilities. These include existing major science facilities such as McMaster Nuclear Reactor [20], Canadian Light Source [21], TRIUMF [22], or those under consideration, such as the Canada Prototype Compact Accelerator-based Neutron Source (PC-CANS) [23, 24]. This is most evident from the lack of irradiation capabilities required for material and fuel testing for the existing fleet or the advanced and future reactors (such as CANDU® MONARK™ Reactor [25], other Generation IV nuclear reactors [26, 27], or upcoming SMRs [13-16]).

### 3.2 Clear support and need for a domestic research reactor

The need for a new domestic research reactor to ensure Canada's energy supply resilience and nuclear safety and security was abundantly clear from the workshop series with the most prominent argument stemming from sovereignty and national security. Canada's status as a Tier 1 nuclear nation is expected to continue with refurbishment of the existing fleet of domestically developed power reactors, and its goal of development and deployment of SMRs (and other advanced reactors) as part of its clean energy mix to achieve net zero by 2050. A new research reactor will enable Canada to achieve these goals and self-reliantly respond to its needs by ensuring access to critical resources necessary in the following key areas:

- Nuclear safety, security, licensing and regulatory, development, testing, and verification of new nuclear technology and tools (such as instruments for monitoring and remote control, cyber security, and obtaining required data in various areas).

- Nuclear energy generation. The new research reactor is needed to support the safe operation, maintenance and life extension of the current/refurbishing fleet, and development of SMRs and other advanced reactors. Materials and fuel irradiations under reactor operating conditions are necessary to demonstrate safe operating margins and to simulate accident behaviours of fuel and fission products, to support regulation, and to develop and qualify materials and fuel for advanced reactors and SMRs (including after they become operational).

- Self-sufficiency in training HQP and required skilled nuclear workers (nuclear operators, engineers, and scientists). The new domestic facility will ensure Canada's self-reliance in training the skilled workers by providing irreplaceable operational experience for the uniquely Canadian technologies.
- Radioisotope production. A domestic facility contributes to Canada's self-reliance for access to critical radioisotopes. An example that highlights the need for self-reliance is radiotracers for regulatory purposes.
- research, development and innovation for industry, academia, and government (material science and engineering, basic and applied sciences)

Moreover, in addition to a shortage of neutron sources and irradiation facilities globally [28, 29] (which is only expected to increase against the backdrop of declining supply), hard to overcome if not impossible challenges in accessing international facilities (including cross-border difficulties when highly radioactive materials such as active fuel or other sensitive specimens are involved and timely access) were discussed in the workshops. A domestic facility will resolve these challenges. It will also avoid foreign dependency ensuring sovereignty is preserved, and the national agenda (e.g., energy, defense, and security) can be prioritized. The new research reactor will help to ensure the protection of Canadian intellectual property. It will also be an excellent source of data for studies of reactor operations, nuclear emissions, and simulated accidents.

It should be noted that while some in the workshops questioned the need, or suggested alternative solutions (such as spallation sources), the dominant theme of discussions was in support of a new research reactor, and discussion often turned to trying to figure how best to make it happen.

### 3.3 Maintaining Canadian leadership in nuclear S&T for decades to come

The rich history of Canada in many areas [30], such as nuclear energy generation, reactor safety, production of radioisotopes, and neutron beam research, was highlighted throughout the workshop series. Considering the significant contributions that Canada has made to all these fields and the desire to retain these capabilities into the future, a new domestic research reactor becomes essential.

The need for a domestic facility to retain the Canadian competency is evident in all its utility. For the neutron scattering community, the impact of NRU shutdown in 2018 is that 40% of users have since not conducted an experiment. The impact of NRU shutdown on materials and fuel irradiation work is even more severe, with no fuel irradiation experiments performed ever since and a reduction in material irradiations except those for commercial customers performed off-shore. While the remaining lower-flux reactors in Canada can compensate for some lost capabilities, the full range of Canada's needs requires a high-flux source.

Without a domestic source, it is difficult to retain the existing expertise and to expand the user base. A domestic facility will support exploratory experiments (for which beam times are difficult to get at foreign facilities) and will provide Canadian researchers and companies priority access to neutron beams for materials research and development (e.g., novel materials with disruptive properties and technological applications).

As new technologies are emerging, a high-flux multipurpose reactor that can adapt and respond to the needs of nuclear industry over the coming decades will be instrumental to position Canada in maintaining its leadership role, and to maximize the benefits to Canada from the development and deployment of these new technologies. A new research reactor will also be key in the development and deployment of transmutation and partitioning technologies, enabling nuclear fuel recycling. Thereby, the research reactor will be central to carbon-free energy.

A new research reactor also provides the state of readiness for Canada in other key areas such as production of emerging radioisotopes with significant promise and future radioisotopes for innovative and critical nuclear diagnosis or nuclear treatment. Moreover, from the discussions throughout the series, it is evident that Canada's role would be much diminished as a world player in nuclear S&T without access to a research reactor. A domestic facility will position Canada to remain at the forefront of all of this cutting-edge research, enabling timely research, as well as attracting new research and talents to Canada.

With a new domestic research reactor, Canada will be able to continue its contributions to the international community and, in return, keep goodwill towards it (e.g., access to international resources for Canadian researchers for exceptional cases where experimental requirements cannot be met by domestic facilities). A new domestic research reactor will also allow Canada to continue with its international obligations and contributions. Examples include Canada's participation in IAEA's programs on safeguards and non-proliferation.

### 3.4 Stakeholder requirements: common priorities & areas of consensus

The presentations and discussions throughout the series indicate that a research reactor with the main characteristics of high flux, wide range and flexibility of neutron spectrum, large irradiation capacity and volume can respond to virtually all the needs of stakeholders. Other specifications and capabilities discussed included: ability to perform both rapid and long-term irradiations, provisions in reactor design for development of instrumentation, neutron beams for a variety of experiments (both thermal and cold) and state of the art neutron beam instruments and specialized sample environments, irradiation sites for radioisotope production and nuclear transmutation, Post Irradiation Examination (PIE) facilities (hot cells, processing facilities, radiochemical laboratories, and other labs for pre-irradiation processing, and flexibility (to address the needs of future power reactors besides those currently under consideration).

On the mission and access policy of the reactor, the inclusion of education, training, and outreach as part of its core mandate was emphasized, and the facility should have an open-access user policy to encourage collaborative R&D across the Canadian nuclear landscape, which would lead to innovation and development of new technologies. This aligns well with the vision of the Chalk River Laboratories as an R&D campus and national laboratory. Thus, other facilities and capabilities that support training in key areas should be included in the research reactor consideration from the start.

### 4. Conclusions

CNL developed and ran a workshop series to gather wide-spread expert opinions for the needs and requirements for a new research reactor in Canada. While some opinions suggested alternative solutions, the dominant theme supported a new high-flux research reactor as alternatives do not fill the gaps and opportunities for a MPRR. From this preliminary stakeholder engagement, several shared understandings have emerged to support the continued development of Canada as a Tier 1 nuclear nation. First, access to existing Canadian and international facilities does not fully and easily meet all the needs and requirements of stakeholders across the Canadian landscape. Second, to ensure that sovereign national priorities are met in key areas of Canada's energy resilience and nuclear safety and security, a new domestic high-flux multipurpose research reactor is required. With such a facility, Canada can self-reliantly respond to its needs and priorities for a continued safe and secure generation and utilization of nuclear power (reactor development, deployment, operations, and related activities) for decades to come. Third, a new domestic high-flux multipurpose research reactor is needed for Canada to remain a global leader in nuclear S&T.

A high-flux multipurpose research reactor will provide the state of readiness, facilitate innovation and future technologies, and attract talent from around the world, all of which contribute to securing the competitive edge of Canadian nuclear S&T. As a focal point, and a complement to the range of university facilities, the new research reactor will bring together all the elements of the Canadian nuclear community (academics, industry, researchers, experts, leaders) and spur numerous opportunities for collaborations across the Canadian landscape. A new research reactor project will also be a driver of jobs and economic growth during its construction, and over decades of service, as a foundation for innovation that contributes substantially to the Canadian S&T landscape (turning science into jobs). It is the only disruptive option that addresses all three pillars of a successful S&T strategy—people, knowledge, and innovation—to ensure Canada's nuclear excellence, competitiveness, and leadership well into the future.

At this early stage of consideration, the workshops provided a forum for interested stakeholders to provide a Pan-Canadian view of the landscape for the need of a research reactor. The workshop series did not attempt to establish a firm and committed stakeholder (customer) base, engage Indigenous and local communities, address how to finance the reactor construction and operation, or identify models that best provide a Canadian solution (including possible partnerships that bring together federal and provincial governments, academia, industry, and international partners). The next steps in the research reactor consideration would include efforts to address these matters.

## 5. Acknowledgements

This work was supported by Atomic Energy of Canada Limited's Federal Nuclear Science & Technology Work Plan. We thank all the presenters and attendees of the workshop series whose participation was essential in gathering insightful considerations for a new research reactor for Canada and the success of the series in reigniting the national dialogue for the reactor. We are also grateful to CNL and AECL for their ongoing support of this initiative.